%% file: main.tex
\documentclass[sigconf, nonacm]{acmart}

\newcommand\vldbdoi{XX.XX/XXX.XX}

\newcommand\vldbavailabilityurl{URL_TO_YOUR_ARTIFACTS}
\newcommand\vldbpagestyle{\vldbpagestyle}

\usepackage{tikz}
\usetikzlibrary{calc}
\usepackage{tikz}
\usepackage{tikz-qtree}
\usetikzlibrary{shapes.geometric, arrows, positioning}
\usepackage{operatorsymbols}
\usepackage{mathtools}
\usepackage{enumitem}

\usepackage{balance}

\usepackage{fdsymbol}

\usepackage{float}

\newcommand{\parachute}{\texttt{parachute}}
\newcommand{\uppara}{\texttt{parachute}$\uparrow$}
\newcommand{\downpara}{\texttt{parachute}$\downarrow$}
\newcommand{\updownpara}{\texttt{parachute}$\updownarrow$}
\newcommand{\PSF}{\textsc{PSF}}
\newcommand{\flow}{\rightsquigarrow}
\newcommand{\MDDL}{\textsc{MDDL}}

\newcommand{\sparagraph}[1]{\vspace{1mm}\noindent {\bf #1}}
\newcommand{\setC}{\mathcal{C}}
\newcommand{\LIKE}{\texttt{LIKE}}
\newcommand{\ILIKE}{\texttt{ILIKE}}

\newcommand{\REGEX}{\texttt{REGEX}}
\newcommand{\pbw}{\texttt{pbw}}
\newcommand{\NULL}{\texttt{NULL}}
\newcommand{\pip}{\texttt{pipeline}}

\newcommand{\tabci}{\texttt{cast\_info}}
\newcommand{\tabmk}{\texttt{movie\_keyword}}
\newcommand{\tabt}{\texttt{title}}
\newcommand{\tabk}{\texttt{keyword}}

\newcommand{\tabit}{\texttt{info\_type}}
\newcommand{\tabmiidx}{\texttt{movie\_info\_idx}}
\newcommand{\tabkt}{\texttt{kind\_type}}

\usepackage[most]{tcolorbox}
\usepackage{amsmath}
\newtcolorbox{example}[1][]{
  enhanced,
  colback=white,         % Background color of main content
  colframe=black,        % Frame color
  coltext=black,         % Text color in main content
  coltitle=black,        % Title text color
  colbacktitle=white,    % Title background color
  boxrule=0.6pt,           % Frame thickness
  arc=1mm,               % Rounded corners
  left=2mm,
  right=2mm,
  top=2mm,
  bottom=2mm,
  fonttitle=\bfseries,   % Title font style
  title=Example,         % Default title
  #1                     % Allow override
}

\definecolor{dkgreen}{rgb}{0,0.6,0}
\definecolor{gray}{rgb}{0.5,0.5,0.5}
\definecolor{mauve}{rgb}{0.58,0,0.82}
\definecolor{mygray}{rgb}{0.5, 0.5, 0.5}
\usepackage{xcolor}
\definecolor{persimmon}{rgb}{0.93, 0.35, 0.0}
\definecolor{darkscarlet}{rgb}{0.34, 0.01, 0.1}
\definecolor{goldenbrown}{rgb}{0.6, 0.4, 0.08}
\definecolor{harvardcrimson}{rgb}{0.79, 0.0, 0.09}
\definecolor{darkviolet}{rgb}{0.58, 0.0, 0.83}
\definecolor{darkpastelgreen}{rgb}{0.02, 0.5, 0.2}
\definecolor{darkpink}{rgb}{0.91, 0.33, 0.5}
\definecolor{mediumslateblue}{rgb}{0.48, 0.41, 0.93}

\usepackage[textsize=tiny]{todonotes}

\newif\ifcolormode
\colormodefalse % Uncomment this line to disable colors

\ifcolormode
  \newcommand{\mr}[1]{{{\color{persimmon}#1}\largertodo[color=persimmon]{MR1}}}
  \newcommand{\mrno}[1]{{{\color{persimmon}#1}}}
\else
  \newcommand{\mr}[1]{{{#1}}}
  \newcommand{\mrno}[1]{{{#1}}}
\fi

\ifcolormode
  
  \newcommand{\mrra}[1]{{{\color{goldenbrown}#1}\largertodo[color=goldenbrown]{MR 2.1}}}
    
  \newcommand{\mrrb}[1]{{{\color{blue}#1}\largertodo[color=blue]{MR 2.2}}}
    \newcommand{\mrrbno}[1]{{{\color{blue}#1}}}
  \newcommand{\mrrc}[1]{{{\color{darkpink}#1}\largertodo[color=darkpink]{MR 2.3}}}
      \newcommand{\mrrcno}[1]{{{\color{darkpink}#1}}}
\else
  
  \newcommand{\mrra}[1]{{{#1}}}
    
  \newcommand{\mrrb}[1]{{{#1}}}
    \newcommand{\mrrbno}[1]{{{#1}}}
  \newcommand{\mrrc}[1]{{{#1}}}
    \newcommand{\mrrcno}[1]{{{#1}}}
\fi

\ifcolormode
  
\else
  
\fi

\ifcolormode
  \newcommand{\mrrrr}[1]{{{\color{darkpastelgreen}#1}\largertodo[color=darkpastelgreen]{MR 3}}}
\else
  \newcommand{\mrrrr}[1]{{{#1}}}
\fi

\ifcolormode
  \newcommand{\opt}[1]{{{\color{mediumslateblue}#1}\largertodo[color=mediumslateblue]{R1.W3}}}
\else
  \newcommand{\opt}[1]{{{#1}}}
\fi

\ifcolormode
  \newcommand{\other}[1]{{{\color{darkscarlet}#1}}}
\else
  \newcommand{\other}[1]{{{#1}}}
\fi

\begin{document}
\title{Parachute: Single-Pass Bi-Directional Information Passing}

%%
%% The "author" command and its associated commands are used to define the authors and their affiliations.

\author{Mihail Stoian}
\orcid{0000-0002-8843-3374}
\affiliation{%
  \institution{University of Technology Nuremberg}
  \streetaddress{}
  \city{Nuremberg}
  \state{Germany}
  \postcode{}
}
\email{mihail.stoian@utn.de}

\author{Andreas Zimmerer}
\orcid{todo}
\affiliation{%
  \institution{University of Technology Nuremberg}
  \streetaddress{}
  \city{Nuremberg}
  \state{Germany}
  \postcode{}
}
\email{andreas.zimmerer@utn.de}

\author{Skander Krid}
\orcid{todo}
\affiliation{%
  \institution{University of Technology Nuremberg}
  \city{Nuremberg}
  \country{Germany}
}
\email{skander.krid@tum.de}

\author{Amadou Latyr Ngom}
\affiliation{%
  \institution{Massachusetts Institute of Technology}
  \streetaddress{}
  \city{Cambridge}
  \state{Massachusetts}
  \postcode{}
}
\email{ngom@mit.edu}

\author{Jialin Ding}
\affiliation{%
  \institution{Amazon Web Services}
  \streetaddress{}
  \city{Boston}
  \state{Massachusetts}
  \postcode{}
}
\email{jialind@amazon.com}

\author{Tim Kraska}
\affiliation{%
  \institution{Massachusetts Institute of Technology}
  \streetaddress{}
  \city{Cambridge}
  \state{Massachusetts}
  \postcode{}
}
\email{kraska@mit.edu}

\author{Andreas Kipf}
\affiliation{%
  \institution{University of Technology Nuremberg}
  \streetaddress{}
  \city{Nuremberg}
  \state{Germany}
  \postcode{}
}
\email{andreas.kipf@utn.de}

% Abstract.
\begin{abstract}
Sideways information passing is a well-known technique for mitigating the impact of large build sides in a database query plan.
As currently implemented in production systems, sideways information passing enables only a \emph{uni-directional} information flow, as opposed to instance-optimal algorithms, such as Yannakakis'. On the other hand, the latter require an additional pass over the input, which hinders adoption in production systems.

In this paper, we make a step towards enabling \emph{single-pass} \emph{bi-directional} information passing during query execution. We achieve this by statically analyzing between which tables the information flow is blocked and by leveraging precomputed join-induced fingerprint columns on FK-tables. On the JOB benchmark, Parachute improves DuckDB v1.2's end-to-end execution time without and with semi-join filtering by 1.54x and 1.24x, respectively, when allowed to use 15\% extra space.
\end{abstract}

\maketitle

% \pagestyle{\vldbpagestyle}
% \begingroup\small\noindent\raggedright\textbf{PVLDB Reference Format:}\\
% \vldbauthors. \vldbtitle. PVLDB, \vldbvolume(\vldbissue): \vldbpages, \vldbyear.\\
% \href{https://doi.org/\vldbdoi}{doi:\vldbdoi}
% \endgroup
% \begingroup
% \renewcommand\thefootnote{}\footnote{\noindent
% This work is licensed under the Creative Commons BY-NC-ND 4.0 International License. Visit \url{https://creativecommons.org/licenses/by-nc-nd/4.0/} to view a copy of this license. For any use beyond those covered by this license, obtain permission by emailing \href{mailto:info@vldb.org}{info@vldb.org}. Copyright is held by the owner/author(s). Publication rights licensed to the VLDB Endowment. \\
% \raggedright Proceedings of the VLDB Endowment, Vol. \vldbvolume, No. \vldbissue\ %
% ISSN 2150-8097. \\
% \href{https://doi.org/\vldbdoi}{doi:\vldbdoi} \\
% }\addtocounter{footnote}{-1}\endgroup

\ifdefempty{\vldbavailabilityurl}{}{
\vspace{.3cm}
\begingroup\small\noindent\raggedright\textbf{PVLDB Artifact Availability:}\\
The source code, data, and/or other artifacts have been made available at \url{https://github.com/utndatasystems/parachute}.
\endgroup
}

\input{sections/introduction}
\input{sections/preliminaries}
\input{sections/parachute}
\input{sections/evaluation}
\input{sections/discussion}
\balance
\input{sections/related-work}
\input{sections/conclusion}

\newpage
\balance
\bibliographystyle{ACM-Reference-Format}
\bibliography{sample}

\end{document}

%% file: sections/introduction.tex
\section{Introduction}\label{sec:introduction}

A database system's task is to answer user queries as fast as possible. As simple as that might sound, getting to the edge of performance is hard and sometimes requires brilliant research ideas to emerge.
One such optimization is pruning tuples and partitions of base tables.
An interesting problem appears when one wishes to do so not only based on base table filters, but also via \emph{join-induced} filters; this is what is known as sideways information passing (SIP)~\cite{sip_1, sip_2}.
The key idea is to pass \emph{information} between tables as an exact or approximate representation of the key space of a joining table.
However, doing this optimally remains a challenge on its own.
What production systems now implement is, if the reader will, a rudimentary application of SIP, namely \emph{probe-side filtering} (PSF): At query execution time, build sides and materializing pipeline sinks will compute, if appropriate, a bloom filter which is passed to the upcoming table scans in the query plan, i.e., in the current or later probe pipelines~\cite{pred_cache, vectorwise}. The main intuition behind this approach is that pre-filtering the tables before the join occurs saves on expensive hash table lookups. This is a valid point, as inaccurate cardinality estimates can lead to incorrect build-side choices, resulting in excessively large hash tables during query execution. As we will argue next, this kind of information passing is \emph{uni-directional}.

\begin{figure}
    \centering
    \includegraphics[width=1.0\linewidth]{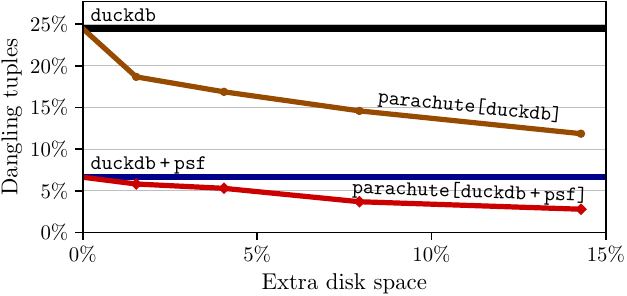}
    \caption{Number of \emph{dangling} tuples in JOB~\cite{job} when running DuckDB v1.2 (\texttt{duckdb})~\cite{duckdb}, probe-side filtering (\texttt{duckdb} + \texttt{psf}), and Parachute on both variants. Dangling tuples are those that do not survive full semi-join reduction and user-provided predicates. The numbers are relative to the total input size, excluding the non-dangling part.}
    \label{fig:fig1}
\end{figure}

In theory, what a system actually \emph{wants} to strive for is a full input reduction, i.e., the filtered base tables eventually contain exactly the tuples that are used in the later join operators. We refer to the other tuples as \emph{dangling} tuples. The type of algorithms capable of achieving zero dangling tuples are instance-optimal algorithms, such as Yannakakis'~\cite{yannakakis} for $\alpha$-acyclic query graphs. Fig.~\ref{fig:fig1} shows the percentage of dangling tuples in JOB~\cite{job} produced by DuckDB~v1.2 (\texttt{duckdb}) and our \PSF{} implementation (\texttt{duckdb\,+\,psf}), described in Sec.~\ref{sec:duckdb-psf}. Note that, for JOB, Yannakakis' algorithm removes all dangling tuples.

However, one of the reasons why instance-optimal algorithms do not make it into production systems is the overhead they incur, as the constant hidden behind the $\mathcal{O}$-notation can be rather large. Indeed, several papers report that Yannakakis' algorithm can be up to 5x slower~\cite{le, pt}.
The underlying issue is that the algorithm itself requires \emph{multiple} passes over the data. This is also true of the more recent approaches, such as Predicate Transfer~\cite{pt}, or its very recent robust version~\cite{rpt}, which aim for an approximate semi-join reduction phase. Since these algorithms enable \emph{bi-directional} information passing, requiring multiple passes seems understandable.

\sparagraph{Research Question.} Inspired by the wide adoption of \PSF{} in production systems, the question is whether it is possible to enable bi-directional information passing through the query plan during query execution, without the need of the additional pass over the input, as required by instance-optimal algorithms.

\sparagraph{Contribution.} We make a step towards answering this question by first formalizing how information is passed in a \PSF-enhanced query plan, showing that it only guarantees a uni-directional information flow, and, secondly, introducing precomputed join-induced fingerprint columns, called \texttt{parachute} columns, that are stored on FK-tables and enabled during query execution time to ensure a \emph{bi-directional} information flow through the query plan. Unlike regular bitmap join indexes~\cite{mt_joins_bji, dimension-join}, \parachute{} columns generalize to high-cardinality columns over non-dimension tables and enjoy a monotonic performance-space tradeoff as shown in Fig.~\ref{fig:fig1}.

Note that Parachute not only supports probe-side pruning, but also build-side pruning if the query plan contains a FK-table on the build side. In a nutshell, Parachute trades off space and insertion performance for improved sideways information passing and hence query performance. We leave it up to the user to control this tradeoff, depending on workload characteristics. In future work, we plan to answer this tradeoff automatically. We show that, on JOB, Parachute achieves a reduction of 8.79x and 2.38x in the total number of dangling tuples over vanilla DuckDB and DuckDB with \PSF{}, respectively. This reduction results in end-to-end execution speedup of 1.54x and 1.24x respectively, with only 15\% space overhead and a 3.9x increase in load time, with further potential for optimization. For CEB, the increase in load time is amortized in a single run, whereas for JOB it takes 15 runs (of 113 queries) to amortize.

In summary, our contributions are as follows:
\begin{enumerate}[noitemsep, topsep=0pt]
    \item We formalize the information flow induced by \PSF{} during query execution.
    \item With Parachute, we achieve \textbf{single-pass} \textbf{bi-directional} information passing by enabling the missing information flow direction in \PSF{}.
    \item We evaluate Parachute on the widely-studied JOB~\cite{job} and CEB~\cite{ceb} benchmarks.
\end{enumerate}

\sparagraph{Organization.} The paper is structured as follows: In Sec.~\ref{sec:preliminaries}, we introduce pipelined execution (Sec.~\ref{subsec:pipelined-execution}), probe-side filtering (Sec.~\ref{subsec:prelim-psf}), and formalize information flow (Sec.~\ref{subsec:information-flow}). In Sec.~\ref{sec:parachute}, we outline Parachute's design, focusing on the way how we represent the join-induced fingerprint columns. We continue with the evaluation on JOB and CEB in Sec.~\ref{sec:evaluation}, and then with a discussion in Sec.~\ref{sec:discussion}. Afterwards, we outline related work (Sec.~\ref{sec:related-work}) and conclude in Sec.~\ref{sec:conclusion}.

%% file: sections/preliminaries.tex
\section{Preliminaries}\label{sec:preliminaries}

In this section, we introduce general notation and formalize pipelined execution (Sec.~\ref{subsec:pipelined-execution}) and probe-side filtering (Sec.~\ref{subsec:prelim-psf}), concluding with the formalization of the information flow (Sec.~\ref{subsec:information-flow}). 

\begin{figure*}
    \centering
    \includegraphics[width=1.0\linewidth]{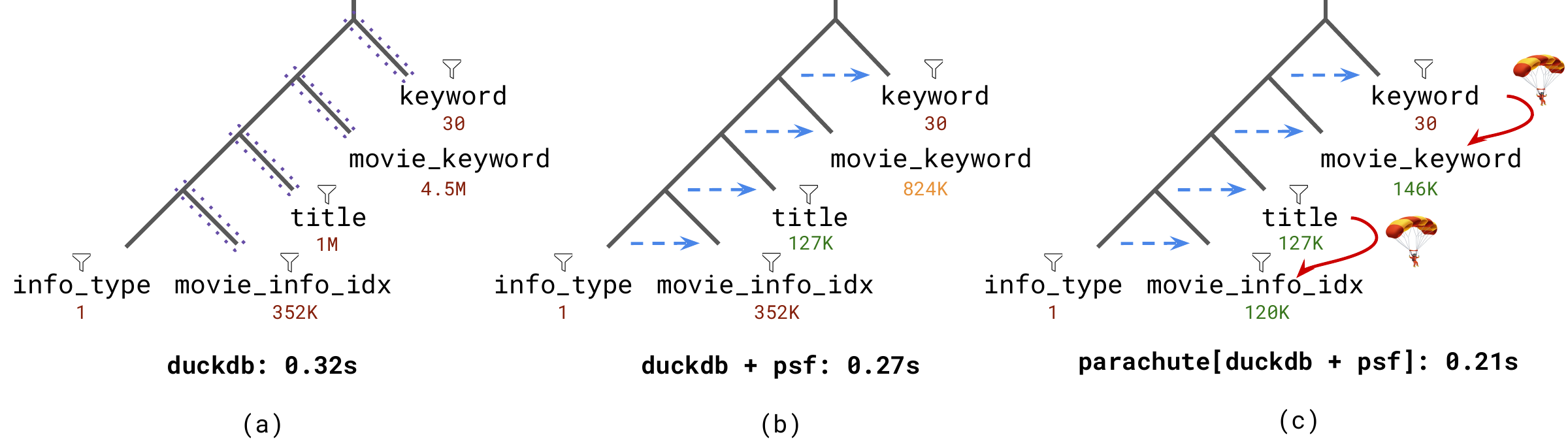}
    \caption{\mrrcno{Visualizing how Parachute activates the downwards information flow in DuckDB's query plan of JOB-4a (v1.2.0, single-threaded). We show (a) the query plan with probe pipelines highlighted, (b) the query plan with the PSF-induced, upwards information flow, and (c) the enabled downwards information flow with \parachute{} predicates.\protect\footnotemark~Note: We follow the convention that the build side is on the left side.}
    \label{fig:full}}
\end{figure*}

\subsection{General Notation}\label{subsec:general-notation}

Given a table (or relation) $T$, we refer to $\mathcal{A}(T)$ as the set of attributes of $T$ (the attributes are grouped in equivalence classes, such that keys like \texttt{movie\_keyword.movie\_id} are equivalent to \texttt{title.id}).

\sparagraph{Query Plan.} A query plan is a binary tree in which the tables lie at leaf nodes, and the join operators at the inner nodes. In the following, we assume a hash-join implementation of the join operator, as this is currently the most efficient, with recent improvements in the last year~\cite{altan_ht, cidr_cwi}. We follow the established convention that the build side is on the left, and the probe side is on the right of the (hash) join operator.

\sparagraph{Semi-Join Reduction.} In the following, by semi-join reduction we mean that a table $S$ can be filtered by $R$, since it holds that $(S\:\lsemijoin\:R) \:\join\:R = S\:\join\:R$. The tuples that have been filtered out from $S$ are called dangling tuples. By an approximate semi-join reducer we mean an approximate representation of the semi-join operator, yet with no false negatives, such that the correctness of the join operator is still guaranteed. This means that $S$ might still contain dangling tuples. A \emph{full} semi-join reduction of a join-query $Q$ is one in which all the tables present in the query have no dangling tuples.

\subsection{Pipelined Execution}\label{subsec:pipelined-execution}

\begin{figure}
    \centering
    \includegraphics[width=1.0\linewidth]{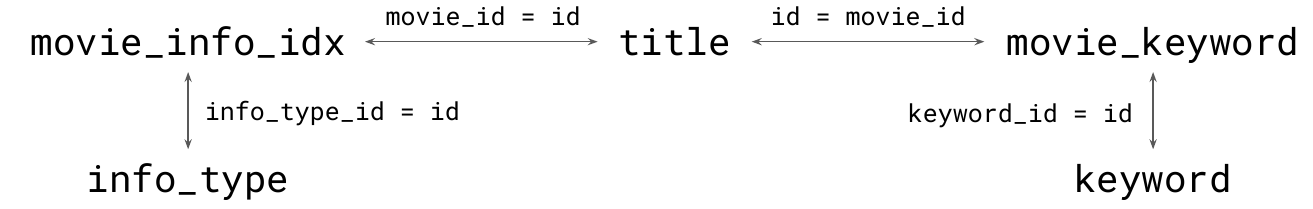}
    \caption{\mrrbno{Query graph of JOB-4a with PK-FK relationships.}}
    \label{fig:query-graph-4a}
\end{figure}

In the following and in the rest of the paper, we assume that query plans are executed via pipelined execution, as currently adopted in state-of-the-art systems~\cite{hyper_pipeline,duckdb,umbra}. In this setting, the query plan is split into multiple execution pipelines. Precisely, a pipeline consists of a probe table and multiple hash tables that are the materializations of other incoming pipelines. To this end, consider \mrrc{the query plan in Fig.~\ref{fig:full}~(a), where we show the query plan of JOB-4a on the IMDb database~\cite{job}}; \mrrb{the query graph of the same query is shown in Fig.~\ref{fig:query-graph-4a}.} To ease the later formalization, we will ignore the (single-table) build pipelines, i.e., we will only consider \emph{probe} pipelines. As a consequence, we uniquely identify a pipeline with its probe table, and say that a pipeline's (single-table) build sides have the same pipeline number as the probe table itself, e.g., \mrra{$\pip(\tabit{}) = \pip(\tabmiidx{})$. With this, the query plan consists of four probe pipelines, which are marked with dotted lines in Fig.~\ref{fig:full}~(a). For instance, in the last pipeline, \pip(\tabk), \tabk{} probes the hash-table built over the output of \pip(\tabmk).}

\sparagraph{Formalization.} This way of executing queries naturally incurs a precedence relation between the (probe) pipelines. Namely, if a pipeline $P_2$ depends on the materialized output of $P_1$ (i.e., $P_1$ needs to execute before $P_2$), then $P_1 < P_2$. \mrra{For instance, in our query plan, we have that $\pip(\tabt) < \pip(\tabmk)$.} While common query plans tend to have this precedence relation as a total one---as is indeed the case with our query plan---this formalization allows for arbitrary ones between pipelines.

\subsection{Probe-Side Filtering (PSF)}\label{subsec:prelim-psf}

It is known that choosing the wrong build side can lead to over-sized hash tables.
Probe-side filtering is a rather pragmatic way employed by production systems to deal with this issue~\cite{pred_cache, vectorwise}: The idea is building bloom filters during \emph{build} pipelines and pushing them into the table scans of subsequent \emph{probe} pipelines to enable early pre-filtering. Hence, \PSF{} mitigates the effect of having to probe large hash-tables. Indeed, consider Fig.~\ref{fig:full}~(b) where we show how \PSF{} pushes bloom filters along the query plan. Note that DuckDB v1.2 already supports a restricted form of PSF: For small tables, such as \tabit{}, it pushes an \texttt{IN}-predicate containing the qualified keys to the probe side. Our implementation of \PSF{} in the same version enables all the other cases (see Sec.~\ref{sec:duckdb-psf} for more details).

\subsection{Information Flow}\label{subsec:information-flow}

In the following, we formalize how information flows in a PSF-enhanced query plan. Intuitively, table $R$ sends information to table $S$ through the query plan if $R$ is on a pipeline that eventually contributes to the build side of $S$'s pipeline, i.e., $\texttt{pipeline}(R) < \texttt{pipeline}(S)$, \emph{and} if $R$ can eventually join with $S$. Here, ``eventually'' refers to the transitive case. To understand this, note that, for instance, \mrra{in the query graph of JOB-4a (Fig.~\ref{fig:query-graph-4a}), \tabit{} cannot \emph{directly} send information to \tabt{}, or vice-versa: it needs \tabmiidx{} to do so. We aim to formalize this in the following.}

\footnotetext{Note that, in Fig.~\ref{fig:full}~(c), for the cardinalities of \texttt{title} and \texttt{keyword}, we ignore the effect of DuckDB v1.2.0's internal optimizations that reduce the cardinalities of these tables further, due to the new \parachute{} predicates.}

\sparagraph{Precedence Relation.} We first define a precedence relation $\prec$ on tables, induced by the pipelined execution. Let $R$ and $S$ be the aforementioned tables. Then,
\begin{equation*}
  R \prec S :\!\Leftrightarrow\\
  \begin{aligned}
    & (\texttt{pipeline}(R) < \texttt{pipeline}(S)) \\
    & \lor (\texttt{pipeline}(R) = \texttt{pipeline}(S) \land \texttt{is\_probe}(S)).
  \end{aligned}
\end{equation*}
Note that we also have to account for the case where $R$ and $S$ lie on the same pipeline, and $S$ is the probe side, in which case $R$ implicitly sends information to $S$ (by a pipeline's definition).

\sparagraph{Is-Joinable.} Next, let $\leftrightarrow$ define the \emph{is-joinable} relation, namely:
\begin{equation*}
    R \leftrightarrow S :\!\Leftrightarrow \mathcal{A}(R) \cap \mathcal{A}(S) \neq \varnothing,
\end{equation*}
where $\mathcal{A}(\cdot)$ is the set of attributes of a given table (Sec.~\ref{subsec:general-notation}). For instance, in the same Fig.~\ref{fig:query-graph-4a}, \mrra{it holds that $\tabmk{} \leftrightarrow \tabk{}$, but also $\tabt{} \leftrightarrow \tabmk{}$.}

\sparagraph{Information Flow.} With this notation in place, we are ready to introduce the information flow relation, which we denote by $\rightsquigarrow$:
\begin{align}
    R \rightsquigarrow S &:\!\Leftrightarrow (R \prec S) \land (R \leftrightarrow S) \land \texttt{is\_probe}(S), \label{eq:base}\\
    R \rightsquigarrow^{n+1} S &:\!\Leftrightarrow \exists T.\; R \rightsquigarrow^n T \rightsquigarrow S, \label{eq:inductive}\\
    R \rightsquigarrow^* S &:\!\Leftrightarrow \exists n > 0.\; R \rightsquigarrow^n S. \label{eq:transitive}
\end{align}
Let us explain the last part in Eq.~\eqref{eq:base}, namely \texttt{is\_probe(S)}: Since \PSF{} only pushes the bloom filters into the probe side, this is a natural addition. Yet, a more advanced \PSF{} could in principle also push to the build sides of upcoming pipelines. In such a case, one can remove this addition. We will see yet another form of SIP, namely LIP~\cite{lip}, and its formalization in Sec.~\ref{subsec:rw-sip}.

Note that the information flow is, by definition, not bound to a single hop. Indeed, we can define its transitive closure $\rightsquigarrow^*$, as above. Intuitively, $R$ passes information to $S$ if there is another $T$ in between such that $R$ passes information to $T$, and $T$ in turn passes information to $S$. Let us see two concrete examples of information flow in the query plan visualized in Fig.~\ref{fig:full}.

\mrra{
\sparagraph{Simple Example.} Consider tables \tabmiidx{} and \tabt{}. It holds that $\pip(\tabmiidx) < \pip(\tabt)$, thus we have $\tabmiidx{} \prec \tabt$. Moreover, since table \tabmiidx{} can join with \tabt{}, i.e., $\tabmiidx{} \leftrightarrow \tabt$, we obtain that $\tabmiidx \flow \tabt$ by Eq.~\eqref{eq:base}.

\sparagraph{Transitive Example.} In the same manner, \tabit{} also sends information to \tabt{}. To see why this is the case, note that \tabit{} is on the same execution pipeline with \tabmiidx{}. As a result, the pushed bloom filter to \tabt{} from the hash-table resulting from this pipeline also contains information from \tabit{}.

Let us see how this works under our formalization. We want to show that $\tabit{} \flow^{2} \tabt$, i.e., two-hop information passing. To this end, we fix $T = \tabmiidx{}$, as specified in Eq.~\eqref{eq:inductive}. We thus have
\[
\begin{array}{c}
    \tabit \flow^{2} \tabt \\
    \iff \\
    \tabit \flow \tabmiidx \land \tabmiidx \flow \tabt.
\end{array}
\]
The second term has already been shown in the previous example, so we are left with showing $\tabit \flow \tabmiidx$. Note that $\tabit{} \prec \tabmiidx{}$ since $\pip(\tabit) = \pip(\tabmiidx)$ and $\tabmiidx$ is the probe side on this pipeline, i.e., $\texttt{is\_probe}(\tabmiidx)$. Finally, we have $\tabit{} \leftrightarrow \tabmiidx$, since \tabmiidx{} has a FK on \tabit{}.
}

\other{
\sparagraph{PSF = Upwards Information Flow.} These examples show us that the information flow induced by PSF is only in the \emph{upward} direction. This naturally limits the query engine’s ability to filter dangling tuples in the opposite, \emph{downward} direction. With Parachute, we will also enable this very direction by leveraging precomputed join-induced fingerprint columns.
}

%% file: sections/parachute.tex
\section{Parachute}\label{sec:parachute}

We have seen that \PSF{} enables an \emph{upwards} information flow. While this can sometimes be enough, and can indeed be achieved in a single pass over the input, with Parachute we make the case that we can also enable the reverse direction, while maintaining the single-pass guarantee. 

\sparagraph{Motivation.} To motivate Parachute, consider again the query plan in Fig.~\ref{fig:full}~(b). Note that \tabt{} has a base-table filter that could have been, in principle, used to filter \tabmiidx{}. However, this is not possible in PSF (while still being single-pass), since \tabt{} is on a probe side. The same is true also for the last probe side in the query plan, namely \tabk{}, which also has a base-table filter. However, in a PSF-enhanced query engine, this cannot be exploited. This is exactly the missing \emph{downwards} information passing which PSF fails to induce. This is where Parachute comes in.

In this case, Parachute recognizes that the information flow is blocked between \tabmiidx{} and \tabt{}, and \tabmk{} and \tabk{}, respectively, and introduces join-induced predicates onto these tables. These are run on auxiliary columns, which are dynamically maintained. The effect of these predicates on the base table sizes is visualized in Fig.~\ref{fig:full}~(c), showing between which pairs of base tables Parachute is active.

\begin{figure}
    \centering
    \includegraphics[width=1.0\linewidth]{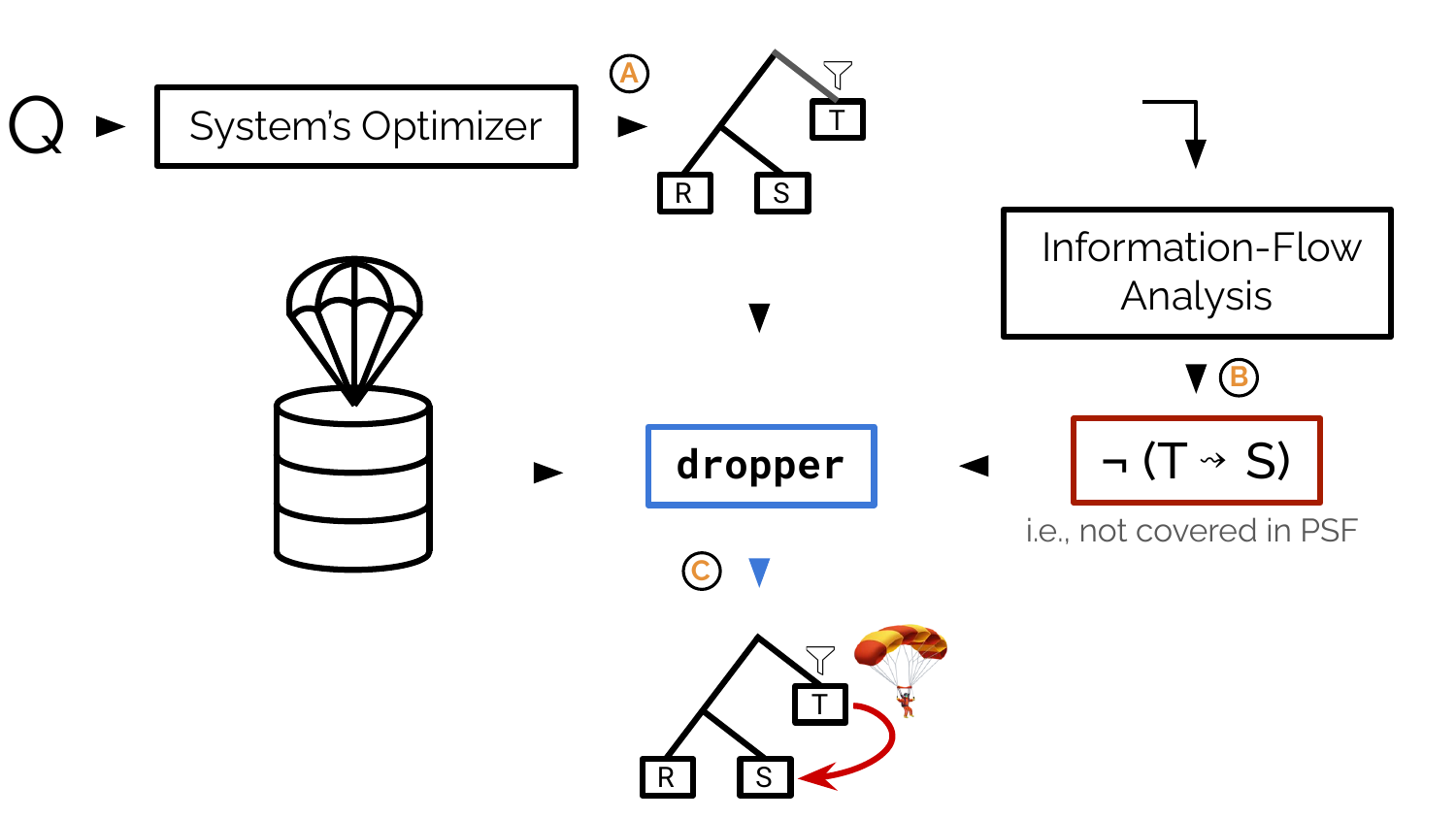}
    \caption{Overview of Parachute's workflow: The system, which is assumed to have \PSF{} enabled, optimizes the query by outputting a query plan (\textcircled{\footnotesize \textcolor{orange}{A}}). Our static information-flow analysis (Sec.~\ref{subsec:information-flow}) identifies the information flow that is not enabled by \PSF{} (\textcircled{\footnotesize \textcolor{orange}{B}}). We use the pre-computed parachute columns in the database to ``drop'' the corresponding parachute predicates on the base tables (\textcircled{\footnotesize \textcolor{orange}{C}}).}
    \label{fig:explainy}
\end{figure}

\sparagraph{Overview.} In Fig.~\ref{fig:explainy}, we overview Parachute's workflow. Query $Q$ is sent to the default system's optimizer, which outputs a query plan (\textcircled{\footnotesize \textcolor{orange}{A}}). System's query engine is responsible to inform Parachute whether it supports any instantiation of SIP, such as PSF (see Sec.~\ref{subsec:rw-sip} for more variants). Then, Parachute runs a static information-flow analysis which outputs a list of \other{ordered table-pairs $(source, target)$ through which information is not flowing in the backwards direction (\textcircled{\footnotesize \textcolor{orange}{B}}). Using this list and the pre-computed and dynamically maintained parachute columns on FK-tables, if the $source$ table---in our case, table $T$---has a base filter, it ``drops'' a parachute predicate on the $target$ table, $S$ (\textcircled{\footnotesize \textcolor{orange}{C}}).} The updated query plan can then be run by the system's engine.

Before we delve into Parachute's design, let us consider a simple example that will motivate the following subsections.

\sparagraph{Example.} Let us assume $Q$ is the following query:
\begin{lstlisting}[label=l:query-example,language=SQL]
select k.keyword
from keyword k, movie_keyword mk, title t
where k.id = mk.keyword_id and mk.movie_id = t.id
and k.keyword like 'Family%'
and t.production_year between 1980 and 2010;
\end{lstlisting}
\mrra{
and set $R \vcentcolon= \tabk{}$, $S \vcentcolon= \tabmk{}$, and $T \vcentcolon=$ \tabt{} in the query plan of Fig.~\ref{fig:explainy}. In this setting, the static information-flow analyzer will inform Parachute that $\tabt{} \not\flow \tabmk{}$. If \tabmk{} has a pre-computed \parachute{} column for \tabt{}'s \texttt{production\_year} column, we can translate the above base filter on \texttt{title.production\_year} in the above query to one adapted for this parachute column. This will be the focus of Sec.~\ref{sec:column-representation}.
}

\subsection{Static Information-Flow Analysis}\label{subsec:analysis}

Our static information-flow analyzer decides between which tables the information flow is not activated in the \emph{downwards} direction. In particular, we will focus on cases such as PK-table $\not\flow$ FK-table pairs, since the \parachute{} columns are present on FK-tables only. Moreover, we can activate a \parachute{} predicate on the FK-table only if the PK-table had one. For instance, in the query plan of JOB-4a (Fig.~\ref{fig:full}), this was the case for the table-pairs (\tabt{}, \tabmiidx{}) and (\tabk{}, \tabmk{}).

The analyzer takes as input the query plan and builds the precedence relation, as explained in Sec.~\ref{subsec:information-flow}. Then, based on schema (or workload) information, it also builds the \emph{is-joinable} relation (recall Sec.~\ref{subsec:information-flow}). Subsequently, it constructs the information flow relation, and its transitive closure $\flow^*$. These constructs can be represented as a binary matrix indexed by the (unique) aliases in the query plan. Finally, it reports the pairs $(source, target)$ for which $source \not \flow^* target$, and $target$ has a FK on $R$.

\subsection{Parachute Column Representation}\label{sec:column-representation}

We now come to the way how we translate columns of arbitrary data types into parachute columns.
Note that, for sake of simplicity, we focus for now on the representation of a parachute column.
We will come to how to ``attach'' this representation to joining tables in Sec.~\ref{subsec:attacher}. In the following, it simply suffices to mention that we target a representation in $\pbw$-many bits, where \pbw{} is the bit-width of the parachute column (independent of the data type of the original column).
The goal is to have a condensed representation of a column $C_i \in \setC$ such that the translation $P'$ of any predicate $P$ on this column does not generate any false negatives; of course, the condensed column should also take less space.
Formally, let $I$ be the set of tuples that qualify for the predicate $P$; similarly, $I'$ for $P'$.
Then, it should hold that $I \subseteq I'$, i.e., no qualifying tuples should be skipped.
This is particularly important when we scan the table that the parachute column is attached to.

In the following, we make the differentiation between numeric- and string-valued columns and show how to support each type.
Indeed, numeric columns are easy to handle, e.g., using histograms.
On the other hand, supporting string-valued columns, in particular \LIKE{} is much more challenging.
We will denote the condensed column as $C_i'$.

\sparagraph{Distribution Estimation.} Before we start, let us understand how we get an estimate for the value distribution of an \emph{attached} parachute column, e.g., \texttt{title.production\_year} on \tabci{}. This will be helpful in modelling the later histograms, which is one possible representation. To this end, we take a random sample of $m = 10^4$ tuples and join it with the PK-table. This helps Parachute understand which values should have, for instance, their own histogram bin, and which can share their bin with other values. This has the ultimate goal of reducing the false positive rate. For example, it is unwise to place two frequent values into the same bin, since when filtering for one, we will not skip the tuples of the other, leading to more (unwanted) false positives. We will thus assume that we have this distribution sample provided for low-cardinality columns. Moreover, values not represented in the sample are assign a default frequency of 1.

\subsubsection{Supporting Numeric Columns}\label{subsubsec:numeric-support}

Assuming we use histograms for numeric columns, let us start with their predicate translation, since this will motivate the \emph{type} of histograms we will be using.

\sparagraph{Numeric Predicate Translation.} Let $P$ be a predicate of type $C_i = a$, where $a$ is a numeric constant.
Using histograms, this can be translated to $C_i' = \texttt{bin}(a)$, where $\texttt{bin}$ maps a value to the corresponding bin.
A similar approach can be applied for range predicates like $C_i\:\texttt{between}\:a\:\texttt{and}\:b$. The corresponding translation of this predicate is $C_i'\:\texttt{between}\:\texttt{bin}(a)\:\texttt{and}\:\texttt{bin}(b)$.
Let us exemplify these translations.

\begin{example}[title=Example: Numeric Predicate Translation]
    Consider the range predicate $\texttt{production\_year}<2025$. If constant 2025 falls into the 7th bin, then the translated predicate reads $\texttt{production\_year}' \leq 7$. Note that we \emph{need} a ``$\leq$'', since the 7th bin may contain years that are less than 2025.
\end{example}

\sparagraph{Value Representation.} The type of histogram we choose has to be responsive to a variety of predicates. To this end, we use \emph{equi-depth} histograms and optimize them directly on the sample using an optimized dynamic programming in $\mathcal{O}(2^{\pbw} m \log m)$-time.

As always, we also need to take into account \NULL{} values (if the respective column $C_i$ is nullable). This can be done by reserving the smallest bin in the histogram for them.

\sparagraph{Numeric UDFs.} Naturally, this histogram representation does not allow for arbitrary predicates, e.g., $\texttt{is\_leap}(\texttt{year})$. Note, however, that if the number of distinct values in the column is small, we can assign each distinct value to a bin and support arbitrary predicates. Indeed, this can still be represented by (equi-depth) histograms: we simply set the bin-boundaries to the distinct values.

\subsubsection{Supporting String Columns}\label{sec:string-fingerprints}

Parachute's design cannot be regarded as realistic if string columns are not supported. The temptation of discarding (or postponing) string support in research is great, however, it is now known that string data is the most prominent in production workloads~\cite{redset}. In the following, we make the distinction between support for low-cardinality and high-cardinality string columns.

\sparagraph{Low-Cardinality Regime.} Assume that we have a low-cardinality string column. If its cardinality is $\leq 2^{\pbw}$, we can assign a histogram bin to each unique value (including \texttt{NULL}, if the column is nullable). When translating a predicate, we simply look up the value in the histogram and take the corresponding bin index. Note that this case is usually covered by SIP~/~PSF, since this corresponds to a predicate on a dimension table.

A more interesting case, still in the low-cardinality regime, is when the cardinality is slightly larger than $2^{\pbw}$; this happens for small $\pbw{}$. We can still apply this technique, yet in an approximate sense. Namely, we build an equi-depth histogram over the value distribution and store a mapping indexing the values to their assigned bins. In this case, a bin will contain more values, yet a frequent one has the chance to be assigned its own bin.

\sparagraph{String UDFs.} By the design of a parachute column, we can indeed support UDFs for the low-cardinality regime. Consider, for instance, the following predicate (assuming a low-cardinality \texttt{title} column, for the sake of the example)~\cite{lotus}:
\begin{lstlisting}[label=l:query-llm,language=SQL]
where LLM((*@\textcolor{mygray}{f}@*)'Is (*@\textcolor{mygray}{\{t.title\}}@*) a boring movie?')
\end{lstlisting}
We can call the LLM on each distinct value, gather the bin indexes corresponding to an affirmative answer, and build an \texttt{IN}-predicate with the set of bin indexes selected.

\sparagraph{High-Cardinality.} We show how to support translations of predicates on string-valued columns.
This is in particular interesting to the research line on partition skipping.
In particular, Parachute supports predicates such as $C_i$ \texttt{like} \texttt{'\%pattern\%'}, and, by extension, \texttt{'\%pattern1\%pattern2\%'}.
Naturally, this also covers prefix- and suffix-based predicates such as \texttt{'\%pattern'} and \texttt{'pattern\%'}, respectively.
The next subsections offer a glimpse on Parachute's support for arbitrary \LIKE{} predicates.

\sparagraph{String Fingerprints.} The main observation is that a pattern has the chance to match a string value if the (multi-)set of characters present in the pattern is a subset of that of the string itself.
To understand this, let us consider the pattern \texttt{'\%utn\%'} and the following string values of a column: \texttt{'utn'}, \texttt{'nutella'}, \texttt{'tone'}.
When computing the sets of characters of each string, we obtain \{\texttt{'n'}, \texttt{'t'}, \texttt{'u'}\}, \{\texttt{'a'}, \texttt{'e'}, \texttt{'l'}, \texttt{'n'}, \texttt{'t'}, \texttt{'u'}\}, and \{\texttt{'e'}, \texttt{'n'}, \texttt{'o'}, \texttt{'t'}\}, respectively.
A first observation is that the first and second strings \emph{could} qualify for the pattern, while the latter cannot, since the character set of the pattern \{\texttt{'u'}, \texttt{'t'}, \texttt{'n'}\} is not a subset of \{\texttt{'t'}, \texttt{'o'}, \texttt{'n'}, \texttt{'e'}\}.
On the other hand, even though the second string value qualifies, it is indeed a false positive. In particular, note that this representation cannot generate false negatives.

The next question is \emph{how} to compactly represent such sets of characters, while supporting arbitrary languages.
If we were to support English only, a 128-sized bitmap per word would suffice: We simply use the ASCII value of a letter and set the corresponding bit to 1.
Before we come to how to reduce the space usage even further, let us first discuss how string predicates are actually translated to this representation.
This will motivate the upcoming paragraphs.

\sparagraph{String Predicate Translation.} Let $P \vcentcolon= C_i\:\LIKE\:\texttt{'\%pattern\%'}$, where \texttt{pattern} is a string constant (recall that this is just a primitive; see the above discussion for the predicate types we support).
We first convert the pattern to its character set representation (without the SQL-specific characters \texttt{'\%'} and \texttt{'\_'}), represented as a 128-sized bitmap \texttt{pmask}. As argued before, a string value has a chance to qualify for this pattern if $C_i'\:\&\:\texttt{pmask} = \texttt{pmask}$, i.e., subset check at fingerprint level. To support \ILIKE{} predicates, build \texttt{pmask} by OR-ing out the masks corresponding to  \texttt{pattern.lower()} and \texttt{pattern.upper()}, respectively.
We will come back with an example when our more compact representation is in place.

\sparagraph{Byte Partitions.} Our simplistic 128-bit bitmap would naturally lead to an increase in the space usage of a parachute column. Indeed, we would require an extra of 16 bytes per tuple.
Ideally, we would have a mechanism to further reduce the space usage of a string parachute column to $\pbw$ bits, where $\pbw$ is the bit-width we also fixed for the numeric ones.
Probably not enough, but do recall that we have to support arbitrary UTF-8 codes as found in current benchmark data, e.g., the IMDb dataset naturally contains names of actors with non-English characters.
Understandably, such codes cannot be covered anymore by our 128-bit bitmap.
Let us first fix this issue.
Note that if we switch to the UTF-8 byte-representation of a string, we are again in a reduced byte space of $[0, 255]$.
Assuming \emph{valid} UTF-8 codes, we can apply the same logic from above on this very byte-representation.
In other words, a byte, even if part of a UTF-8 code, is treated as a standalone byte that can be added to the fingerprint.
This allows to work solely within the byte space $[0, 255]$.
Next, we fix the space usage issue.

We argue that partitioning the byte space into cleverly-selected $\pbw$-many clusters solves our problem.
Namely, a byte is assigned to a unique cluster.
When fingerprinting a string, we (a) iterate its UTF-8 byte-representation, (b) lookup its cluster, (c) set a 1 to the corresponding cluster-bin.
Let us exemplify this construction.

\begin{figure}[h]
    \centering
    \includegraphics[width=0.75\linewidth]{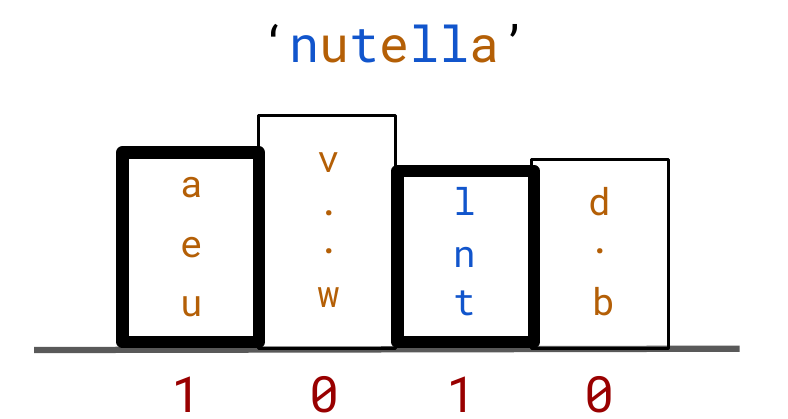}
    \caption{Parachute's string fingerprints: A string value is transformed into its corresponding fingerprint. The byte space---such as the ASCII table---is clustered into four bins, corresponding to four \emph{disjoint} byte clusters.}
    \label{fig:nutella}
\end{figure}

\sparagraph{Example.} Consider the example illustrated in Fig.~\ref{fig:nutella} for the string value \texttt{'nutella'}. In the case of $\pbw = 4$, there are four bins available, corresponding to the clusters which the $[0, 255]$ byte space has been partitioned into. The set of letters appearing in this word are $\{$\texttt{'a'}, \texttt{'e'}, \texttt{'l'}, \texttt{'n'}, \texttt{'t'}, \texttt{'u'}$\}$. The letters \texttt{'a'}, \texttt{'e'}, and \texttt{'u'} fall into the first cluster, while the others in the third cluster. Hence, the final fingerprint reads \texttt{1010}.

The previous example already highlights an aspect that we have not yet optimized for:
We have to ensure that a fingerprint does not have too many ones, otherwise the pattern mask \texttt{pmask} would not skip any tuples. This leads us to a rather intriguing optimization problem that we entitle \textsc{BytePartitioning}, defined below.

% Problem definition.
\begin{definition}[\textsc{BytePartitioning}]
    Given a list of $[0, 255]$-valued strings and a parameter $k$,
    partition the byte space into $k$ clusters such that the number of ones in the fingerprints is minimized.
\end{definition}

We leave it as future work to build these partitions optimally. Currently, we assume a uniform distribution over the bytes, i.e., we use a Round-Robin strategy to build the partitions.

\sparagraph{\REGEX{}-Predicates.} We can also support a limited number of regex classes via this design. Consider the following example:
\begin{lstlisting}[
    label=l:query-regex,
    language=SQL
]
where t.title ~ 'house(keeping|work)?'.
\end{lstlisting}
In this (admittedly very restrictive) setting, we can resort to an \texttt{IN}-predicate with three values, namely \texttt{'house'}, \texttt{'housekeeping'}, and \texttt{'housework'}. Hence, such regular expressions can be translated by enumerating the induced string values.

\subsection{Attaching to Joining Tables}\label{subsec:attacher}

So far we only discussed how to \emph{represent} parachute columns, at least at a conceptual level -- notably, we did not discuss how these are stored and, probably more important, \emph{where}.

\begin{figure}
    \centering
    \includegraphics[width=1.0\linewidth]{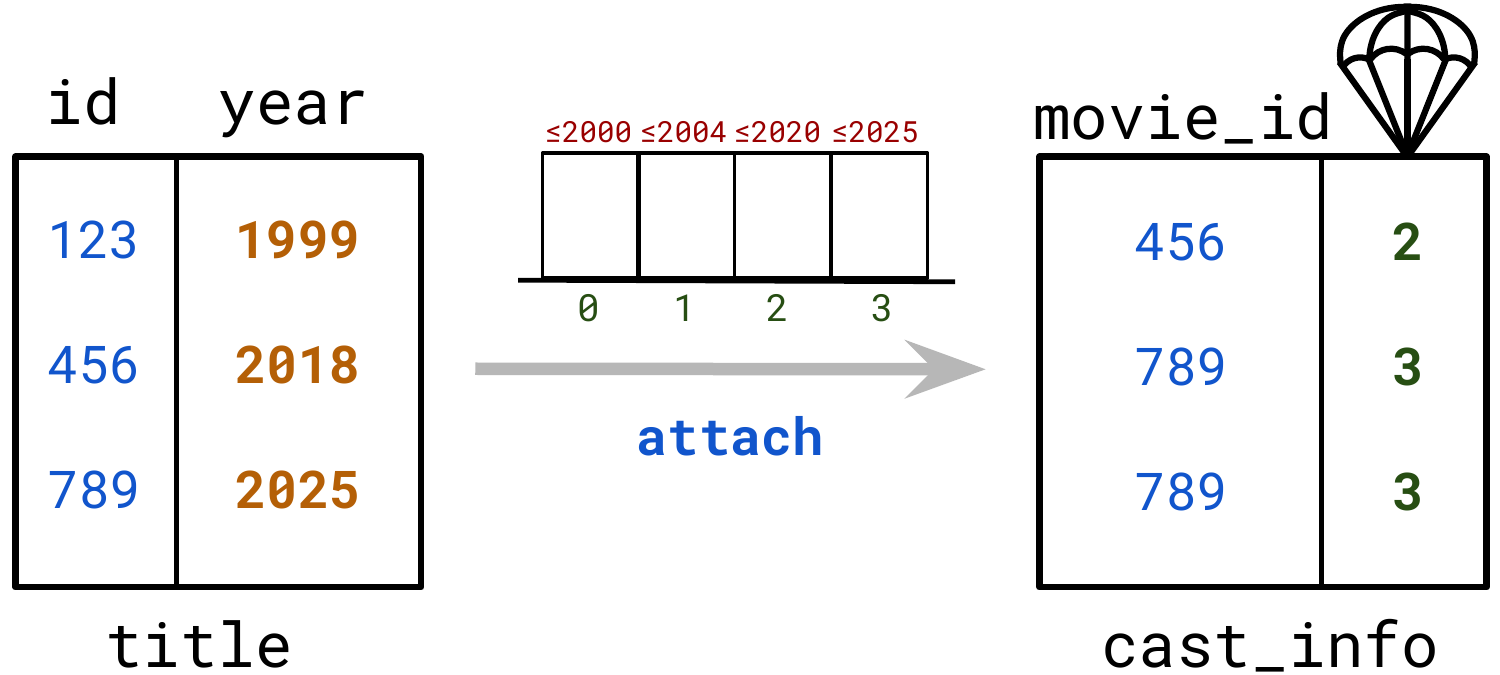}
    \caption{Attaching \tabt.\texttt{production\_year} to \tabci. In this case, we have a parachute bit-width of two, hence four (equi-depth) histogram bins over the domain.}
    \label{fig:attach-vis}
\end{figure}

Indeed, Parachute stores its join-induced fingerprints on FK-tables. Let us take a simple example from the IMDb schema~\cite{job}, namely, attaching a parachute on \tabci{} from \tabt{} on its \texttt{production\_year} column, visualized in Fig.~\ref{fig:attach-vis}. Table \tabci{} stores the casting information and has a FK from \tabt{}, which stores the actual movies. Since \texttt{production\_year} is a numeric column, we will build according to Sec.~\ref{subsubsec:numeric-support} an equi-depth histogram for it. As each tuple from \tabci{} corresponds to exactly one tuple from \tabt{}, we can encode the histogram bin in \pbw{} bits.

\sparagraph{Example.} Concretely, in the example of Fig.~\ref{fig:attach-vis}, the first tuple from \tabci{} has \texttt{movie\_id} = 456, so it will take the \texttt{production\_year} of the second tuple in \tabt{}. Its value is mapped to the third bin in the histogram (hence, 2 in 0-based notation), and set in \texttt{cast\_info}. 

Note that we assumed that PK-FK-constraints are preserved. We show how to allow for looser restrictions, as found in today's data warehouses such as Redshift, in Sec.~\ref{sec:discussion}.

\sparagraph{General Case.} In fact, this very example already showcases how we deal with attaching the parachute column of $C_i$ to FK-tables. We can perform a join between the two tables, during which we introduce a new column that represents the corresponding value of the condensed column $C_i'$. For instance, the \texttt{attach} operation of \texttt{title.production\_year} onto \texttt{cast\_info} reads as follows:
\begin{lstlisting}[
    label=l:query-attach,
    language=SQL
]
update cast_info
set cast_info.parachute_title_production_year = (
  case
    when title.production_year <= 2000 then 0
    when title.production_year <= 2004 then 1
    when title.production_year <= 2020 then 2
    else 3
  end)
where title.id = cast_info.movie_id;        
\end{lstlisting}
Note that if we have to attach multiple parachute columns at once, we will do so in the same \texttt{UPDATE} statement.

This construction readily applies to both numeric- and string-valued columns. To slightly boost this construction for strings, note that we rely on ``helper'' columns: We first build the parachute "locally", i.e., on the base table, and then attach it to the FK-table as is. This saves time due to the fact that we do not build the string fingerprints on the larger table (the FK-table), but rather on the smaller one.

\subsubsection{Transitivity}

A feature of Parachute is that it can be extended to go beyond neighboring, joining tables, and even on \emph{different} join keys. This indeed happens in the IMDb schema, where we have a PK-chain going from \tabci{} $\rightarrow$ \tabt{} $\rightarrow$ \tabkt{}. In such cases, we can have a \parachute{} column attached to \tabci{} which comes from \tabkt{}. To support this in the build phase, we have to perform a topological sort on the schema-induced DAG, such that the parachute column on \tabt{} is already available when building for \tabci{}. Hence, the \texttt{attach} operation may consider already present parachutes on the PK-table. However, since such PK-chains start at dimension tables, which are intrinsically covered by \PSF{}, such parachute columns never get to be used.

\subsection{Inserts and Updates}

We outline Parachute's support for inserts and updates.

\sparagraph{Inserts.} If a batch of new data $B$ is inserted into the FK-table, we need to compute the parachute column only for $B$. To this end, we perform a join of $B$ with the PK-table to attach the respective parachute column on the new slice of the table. We evaluate this overhead in Sec.~\ref{subsec:insert-eval}. Note that since the join intrinsically performs histogram lookups, in case the histogram is not updated, we will map the values larger than the maximum histogram boundary into to the last bin; similarly for the smaller values than the minimum histogram boundary. In case the histogram lost its equi-depth guarantee by a given threshold, we only need to re-attach, hence re-compute, the corresponding parachute column on the FK-table.
In case an insert happens on the PK-table, this is a noop.

\sparagraph{Updates.} The more interesting case are updates, since these may happen on the PK-table. Let us assume that we modified the title of a movie in table \tabt{}, and there is a corresponding parachute column on \tabci{}. Then, we have to re-compute for the join-partners in \tabci{} the parachute values.

%% file: sections/evaluation.tex
\begin{figure*}
    \centering
    \includegraphics[width=1.0\linewidth]{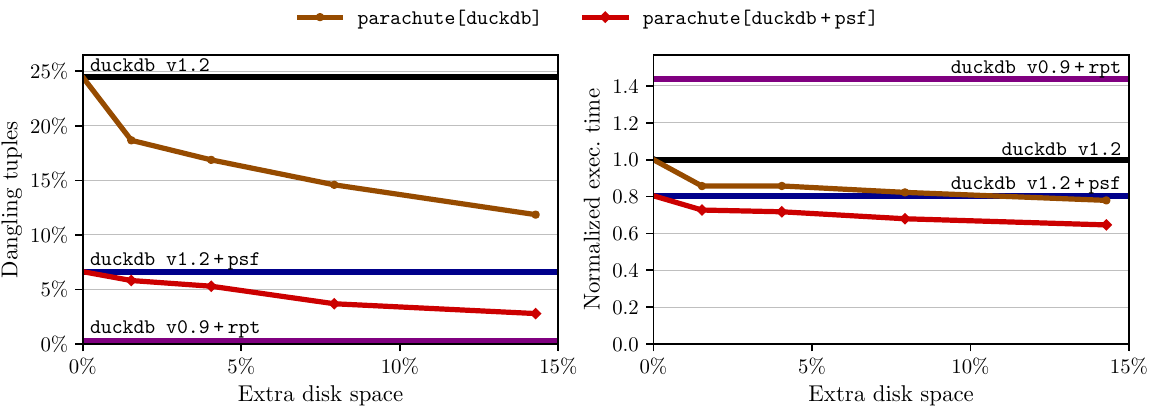}
    \caption{\mrno{The effect of using Parachute on JOB~\cite{job}. We compare with DuckDB v1.2 (\texttt{duckdb v1.2}), our PSF implementation in the same version (\texttt{duckdb v1.2 + psf}), and the recent RPT (\texttt{duckdb v0.9 + rpt})~\cite{rpt}. While \texttt{parachute[duckdb]} enables both the up- and downwards ($\updownarrow$) information flow, \texttt{parachute[duckdb\,+\,psf]} enables only the missing downwards ($\downarrow$) direction, since PSF already guarantees the upwards one (cf.~Sec.~\ref{subsec:information-flow}).}}
    \label{fig:job-plot}
\end{figure*}

\section{Evaluation}\label{sec:evaluation}

\sparagraph{Setup.}~We conduct the experiments on a single node Intel\textsuperscript{\textregistered} Xeon\textsuperscript{\textregistered} Gold 5318Y CPU (24 cores, 48 hyper-threads). The machine is equipped with 128GB DDR4 main memory and runs Ubuntu 24.04. All experiments are run single-threaded due to DuckDB v1.2's limited supported for multi-threaded \texttt{UPDATE} statements.

\sparagraph{Benchmarks.} We evaluate on the well-known JOB~\cite{job} and CEB~\cite{ceb} benchmarks. JOB has 113 queries of 33 templates on the IMDb database, while CEB has a total of 13,646 queries on the same database, yet with a much more variety of base table filters. All benchmarks are run on hot data, following the guidelines in Ref.~\cite{bench_pitfalls}.

\sparagraph{Competitors.} Our PSF implementation (\texttt{duckdb\,+\,psf}) is done on DuckDB v1.2.0 (\texttt{duckdb}) and is described thoroughly in Sec.~\ref{sec:duckdb-psf}. \mr{Moreover, we also consider the recent Robust Predicate Transfer (\texttt{duckdb\,+\,rpt})~\cite{rpt}, which has been implemented on top of DuckDB v0.9.2.\footnote{\texttt{https://github.com/embryo-labs/Robust-Predicate-Transfer}} Therefore, for a complete comparison, we also run Parachute on DuckDB v0.9.2 when comparing with RPT.\footnote{We are aware of an ongoing work to integrate RPT in v1.2.} Note that, starting with v1.2, DuckDB indeed adopted a lightweight SIP implementation, as described in Sec.~\ref{sec:duckdb-psf}. For brevity, we will omit version numbers when the context makes them clear.}

Let us describe how Parachute is integrated with \texttt{duckdb} and \texttt{duckdb\,+\,psf}: Each \parachute{} column is a non-nullable integer column of up to \pbw{} bits. String fingerprints are computed using \texttt{duckdb}'s support for user-defined Python functions. To keep the query plans unchanged if a parachute predicate occurs in the SQL query, we modified \texttt{duckdb} to ignore cardinality estimates for base tables and non-pushed base-table filters. Then, a Python framework maintains a catalog with the existing parachute columns and, based on the result of the static analyzer (Sec.~\ref{subsec:analysis}), drops the corresponding \parachute{} predicates in the SQL query text. We set $\pbw \in \{2, 4, 8, 16\}$ and consider three types of applying Parachute: \uppara{}, \downpara{}, and \updownpara{}, which corresponds to enabling the upwards, downwards, and bi-directional information flow, respectively. We only use one-hop parachute columns in the experiments, i.e., no transitive parachutes.

\input{extra/sip}

\begin{figure*}
    \centering
    \includegraphics[width=1.0\linewidth]{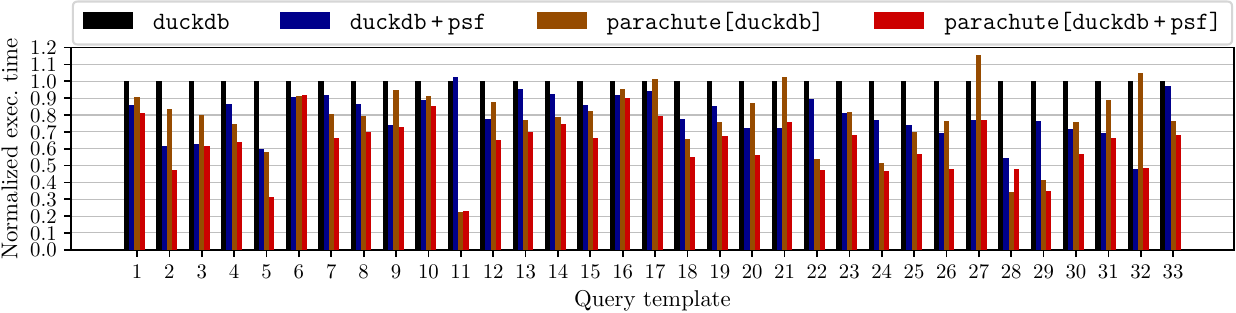}
    \caption{\other{JOB: Normalized execution time in DuckDB v1.2 across query templates. We use a \parachute{} bit-width (\pbw{}) of 16 bits.}}
    \label{fig:job-all}
\end{figure*}

\begin{figure}
    \centering
    \includegraphics[width=1.0\linewidth]{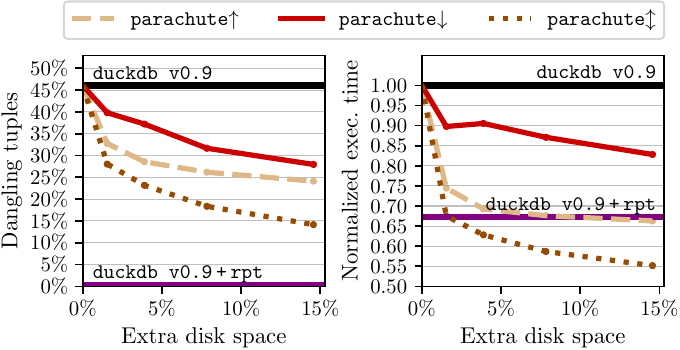}
    \caption{\mrno{JOB: Evaluation of RPT~\cite{rpt} and different Parachute configurations ($\uparrow$, $\downarrow$, $\updownarrow$) on DuckDB v0.9 (\texttt{duckdb v0.9})}.}
    \label{fig:rpt-plot}
\end{figure}

\subsection{JOB \& CEB Results}

In principle, Parachute can infer from the schema which columns can be attached. However, using workload information, one can restrict the number of columns used. Given that query repetitiveness in real workloads is high~\cite{redset}, one can enable Parachute once a set of query templates is already available; we will continue this discussion in Sec.~\ref{sec:discussion}. To this end, we analyze JOB's and CEB's queries and build parachutes only for columns that are also filtered in combination with the respective table.

\sparagraph{JOB.} We show the number of dangling tuples (relative to total input size, excluding the non-dangling part) and the normalized execution time for the entire JOB in Fig.~\ref{fig:job-plot} (see Fig.~\ref{fig:rpt-plot} for an evaluation on v0.9). Let us explain what the competitors stand for: \texttt{parachute[duckdb]} refers to $\parachute\!\updownarrow$ (without \PSF{} activated) since DuckDB v1.2 only implements a rather simplistic PSF, hence the upwards information flow is rather scarce, while \texttt{parachute[duckdb\,+\,psf]} refers to $\parachute\!\downarrow$, since \PSF{} already guarantees the upwards information flow.

The first observation is that we obtain a monotonic behavior of the number of dangling tuples with increasing \pbw{}. In particular, \texttt{parachute[duckdb\,+\,psf]} reaches a number of dangling tuples of 2.79\%. In constrast, vanilla PSF achieves only 6.63\%. The same holds for the normalized execution time (Fig.~\ref{fig:job-plot}, right), where \texttt{parachute[duckdb\,+\,psf]} achieves a speedup of 1.54x and 1.24x over \texttt{duckdb} and \texttt{duckdb\,+\,psf}, respectively, for 14.35\% extra space, or 1.47x and 1.18x, respectively, for 7.94\% extra space.

\mr{
\sparagraph{Parachute vs.~RPT.} Notably, due to its forward and backward passes of the bloom filters, RPT (\texttt{duckdb v0.9\,+\,rpt}) remains with the fewest number of dangling tuples (0.29\%). The reason is that a \parachute{} column has a higher false positive rate compared to a bloom filter due to its narrow bit-width (\pbw{}). For a fair comparison, we show Parachute's effect on DuckDB v0.9.2 in Fig.~\ref{fig:rpt-plot}. RPT indeed improves over vanilla DuckDB by 1.48x, while Parachute achieves a 1.81x speedup when allowed to use 16-bit \parachute{} columns.
}

\sparagraph{JOB Templates.} In Fig.~\ref{fig:job-all}, we show the normalized execution across query templates in JOB with respect to DuckDB v1.2 (\texttt{duckdb}). Each query template's execution time is the sum of the execution times of its instantiations. We see that \texttt{parachute} alone is not sufficient to achieve the best performance: it needs the upwards pass guaranteed by PSF. The regressions in \texttt{parachute[duckdb]}, e.g., \other{JOB-27*}, are due to DuckDB's current limited support to push complex predicates, such as \texttt{col\,\&\,pmask\,=\,pmask}, into the table scan, forcing a materialization of the filtered parachute columns into the projection operator. Recall that such predicates are coming from translated \LIKE{} predicates on high-cardinality columns (Sec.~\ref{sec:string-fingerprints}).
 
\begin{figure}
    \centering
    \includegraphics[width=1.0\linewidth]{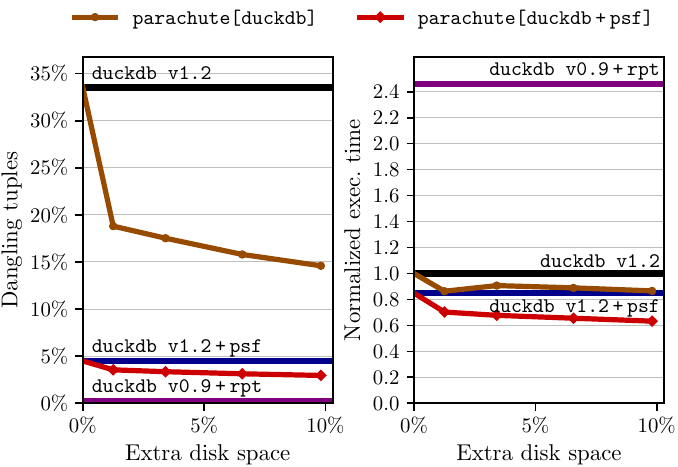}
    \caption{The effect of enabling the downwards information flow in probe-side filtering (\texttt{duckdb\,+\,psf}) on CEB~\cite{ceb}.}
    \label{fig:ceb-plot}
\end{figure}

\sparagraph{CEB.} The same trend can be observed in CEB, the results of which are shown in Fig.~\ref{fig:ceb-plot}. Since there are less parachute columns to maintain (see the next paragraph for the exact numbers for JOB and CEB), we can achieve the same numbers with less extra space: With 9.82\% extra space, \texttt{parachute[duckdb\,+\,psf]} reaches 2.96\% dangling tuples, and a speedup of 1.56x and 1.33x over \texttt{duckdb} and \texttt{duckdb\,+\,psf}, respectively. \mr{In contrast, RPT's strategy removes almost all dangling tuples on this workload (0.18\%).}

\input{data/load-time}

\sparagraph{Size \& Build Time.} Naturally, computing parachute columns incurs some overhead. We report the total time to compute them, for various values of \pbw{} $\in \{2, 4, 8, 16\}$ for both JOB and CEB in Tab.~\ref{tab:workload-size}. For comparison, we also report the time is takes to load the entire IMDb database into DuckDB v1.2 (single-threaded, to ensure a fair comparison). Note that classical FK-indexes on all FK-tables would require an extra space of 62.03\%, assuming no index overhead and 8-byte TIDs.

\begin{figure}[!t]
    \centering
    \includegraphics[width=1.0\linewidth]{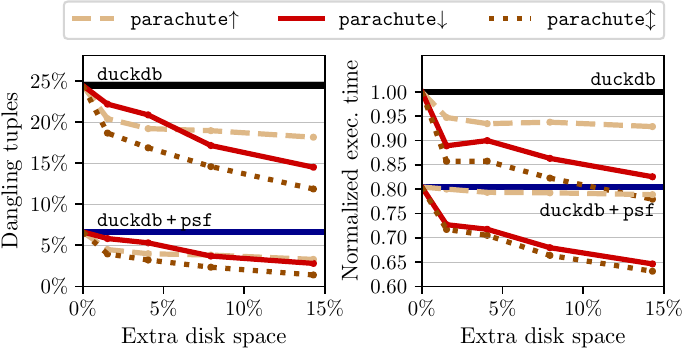}
    \caption{Analyzing different information flow directions ($\uparrow$, $\downarrow$, $\updownarrow$) for Parachute on \texttt{duckdb} and \texttt{duckdb\,+\,psf} in JOB.}
    \label{fig:job-micro-bench}
\end{figure}

\subsection{Micro-Benchmark}

Next, we take Parachute under the lens to understand whether our theoretical formalization is reflected in practice. Namely, we want to answer the following two questions:

\begin{itemize}
    \item[\textbf{Q1}.] Is PSF enough to guarantee upwards information flow?
    \item[\textbf{Q2}.] Is downwards information flow what we should strive for?
\end{itemize}

To this end, consider Fig.~\ref{fig:job-micro-bench}, in which we enable Parachute in different scenarios: The symbol $\uparrow$ represents the upwards information flow (as in PSF), while $\downarrow$ represents the downwards information flow (the one missing in PSF), and $\updownarrow$ the activation of both directions. 

\sparagraph{Q1.} We observe that \uppara{}, while it still decreases its number of dangling tuples, it \emph{cannot} compete with PSF, due to the low false-positive rate achieved by bloom filters. This is also reflected in the plot of the normalized execution time, where it plateaus after a certain point.

\sparagraph{Q2.} This question is more interesting, since it also implies the question whether enabling \emph{both} directions for Parachute, even in the presence of \PSF{}, may provide a much better pruning rate, in this case 1.38\% remaining dangling tuples for \pbw{} = 16. In part, this is natural, since Parachute may indeed help the bloom filter further prune dangling tuples that turn up to be false positives. However, if this is not the case, then the additional parachute predicates only slow down the query execution. In this particular case, the difference in terms of execution time between \downpara\texttt{[duckdb\,+\,psf]} and \updownpara\texttt{[duckdb\,+\,psf]} is minimal, showing that, indeed, the downward direction is what we should eventually optimize for.

\subsection{Inserts}\label{subsec:insert-eval}

\begin{figure}
    \centering
    \includegraphics[width=1.00\linewidth]{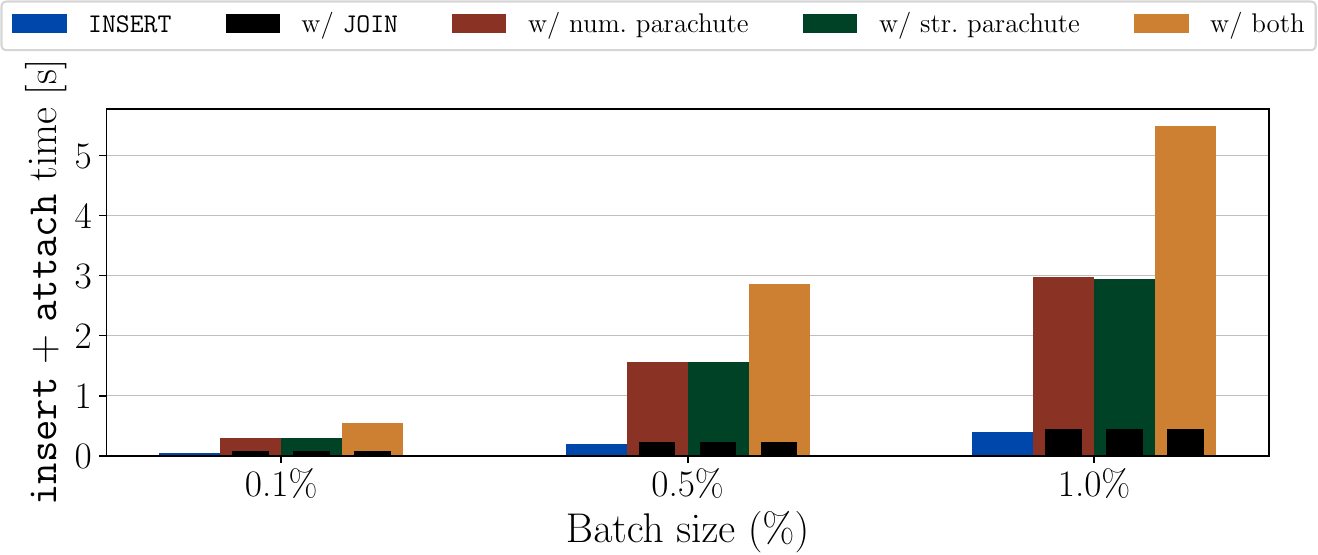}
    \caption{Inserts: We insert batches of various size percentages $\in \{0.1\%, 0.5\%, 1\%\}$ of the original table size, in this case, \tabci{} (36M tuples). We benchmark on the Parachute-enhanced database with \pbw{} = 4 and string helper columns. We stress that DuckDB v1.2 performs \texttt{UPDATE} statements single-threaded.}
    \label{fig:insert-plot}
\end{figure}

Recently, researchers from industry urged the database community to consider how inserts and updates affect the proposed solutions~\cite{redset}. Since Parachute maintains its columns on FK-tables, an insert is simply a lookup by key in the respective PK-table (which, in the absence of an index, can be done by fetching first the corresponding block and then scanning it). We showcase this overhead for batched inserts in Fig.~\ref{fig:insert-plot}, where we insert batches of various size percentages $\in \{0.1\%, 0.5\%, 1\%\}$ into IMDb's \tabci{} (36M tuples). The referenced table in this case is \tabt{}, and the parachute columns are on \texttt{production\_year} and \texttt{title} for the numeric and string case, respectively. Moreover, for the string case, we use the string-helper column maintained on the PK-side. We see that, indeed, this construction blends the difference between numeric and string columns.

Since DuckDB v1.2 runs \texttt{UPDATE} statements single-threaded---enabling multi-threading even slows them down, we also show how much the actual join between the \emph{new} added batch and \tabt{} actually incurs. Indeed, we see that the overhead comes from the update itself.

%% file: extra/sip.tex
% Comes in Evaluation::Competitors.
\subsection{Sideways Information Passing in DuckDB}\label{sec:duckdb-psf}

DuckDB v1.2~\cite{duckdb} implements some form of fundamental sideways-information-passing (SIP) for filtering tuples on the probe sides of hash-joins (PSF).
In the general case, DuckDB transfers the min/max values of the key column(s) to the probe side and uses these values for coarse-grained pruning and filtering.
Additionally, two forms of SIP are only applied only if the number of distinct join-keys on the build side is small.
More specifically, if there is only one value on the build side, an equality predicate on the key-column(s) of the join is generated and transferred to the probe side's base table.
Similarly, if the number of distinct keys on the build side is below a certain threshold (default: 50), an \texttt{IN}-list filter is pushed to the probe side's base table for partition-level pruning, but is not used for row-level filtering.
For build sides exceeding the threshold for the number of distinct keys, only min/max transfer is used.\footnote{We refer the interested reader to the references in the original blog post: \texttt{https://duckdb.org/2025/02/05/announcing-duckdb-120.html\#optimizations}.}

To establish a state-of-the-art baseline for PSF, we implemented bloom joins~\cite{mackert1986r, mullin1990optimal} into DuckDB to perform probe-side filtering on a row-level.
Our implementation provides an end-to-end speedup of 1.26x over vanilla DuckDB on the JOB benchmark.

\sparagraph{Implementation.}
We implemented a custom bloom filter that is tightly integrated into DuckDB's query processing framework.
To guarantee fast builds and lookups \emph{without regressions}, the bloom filter's size is capped at 8\,KiB, ensuring it fits into the processors L1 cache.
Further, we use the 64-bit hash values from the join's hash table to generate two bit positions in the bloom filter that should be set by splitting them into their upper and lower 32 bit and applying fast modulo reduction~\cite{fastmod} with respect to the bloom filter's size.
This design corresponds to a bloom filter with $m=2^{16}$ and $k=2$, supporting about $5000$ distinct keys with a $2\%$ false-positive rate (\textit{fpr}). To maintain the \textit{fpr}, the filter is discarded after construction if more than $34\%$ of bits are set.
During construction, hash values from the hash table are reused, avoiding extra hash calculations.
On the probe side, the bloom filter is evaluated inside the table scan operator directly after simple table filters.
We chose to evaluate the bloom filter after table filters based on the assumption that table filters are faster to evaluate.
If the Bloom filter is insufficiently selective, i.e., filtering less than $60\%$ of rows after $4000$ processed rows, it is automatically disabled.

%% file: data/load-time.tex
\begin{table}
    \centering
    \caption{Building \parachute{} columns on the IMDb database for JOB and CEB: We report for each parachute bit-width $\pbw{} \in \{2, 4, 8, 16\}$ how much time it takes to compute the parachute columns for the entire database and the extra space needed. Note that DuckDB v1.2 performs \texttt{UPDATE} statements single-threaded.}
    \begin{tabular}{lccccc}
        \toprule
        \multicolumn{6}{c}{IMDb \texttt{duckdb}'s \texttt{load}: 91.67s} \\
        \midrule
        & \multicolumn{2}{c}{JOB: 32 parachutes} & \phantom{a} & \multicolumn{2}{c}{CEB: 20 parachutes} \\
        \cmidrule(lr){2-3} \cmidrule(lr){5-6}
        \pbw{} & Attach time & Extra space & & Attach time & Extra space \\
        \midrule
        2  & 244.70s  & +1.53\% & & 184.38s  & +1.25\% \\
        4  & 248.44s  & +4.03\% & & 187.11s  & +3.42\% \\
        8  & 247.61s  & +7.94\% & & 188.47s  & +6.55\% \\
        16 & 265.85s  & +14.35\% & & 190.18s  & +9.82\% \\
        \bottomrule
    \end{tabular}
    \label{tab:workload-size}
\end{table}

%% file: sections/discussion.tex
\section{Discussion \& Future Work}\label{sec:discussion}

\sparagraph{Is Extra Space an Issue?} One could argue that the extra space required by Parachute to store the parachute columns would impede its adoption in production OLAP systems. This may seem understandable at first glance, but we are actually witnessing a trend in OLAP systems to ignore the storage footprint due to its low cost. For example, Amazon Redshift recently added support for MDDL~\cite{mddl} as a new paradigm for sorting the table based on workload predicates. This is well motivated by the high query repetition rate seen in real customer workloads. To this end, MDDL maintains a \emph{new} wide bitmask column, where each bit corresponds to a predicate on the base table. Hence, we do indeed see the openness of an OLAP system towards accepting new columns in its table files.

\sparagraph{Parachute in the Cloud.} There are many ways to prepare Parachute for the cloud. Notably, it comes with a clear advantage over pushed bloom filters: Since a parachute column essentially acts as a ``built-in'' bloom filter into the table, i.e., the parachute column is already attached to the table to be pre-filtered, one can exploit this as part of the \emph{partition pruning} phase using zonemaps~\cite{zonemaps}. While this naturally works for numeric columns, we can apply the same strategy for string fingerprints as well (Sec.~\ref{sec:string-fingerprints}), e.g., by storing min/max-zonemaps and applying \MDDL{}'s skipping technique on its bitmask-columns (see Ref.~\cite[Lst.~3]{mddl}).

\mrrrr{
\sparagraph{PK-FK Relationships.} One challenge that Parachute will need to address in the future is the lack of support for PK-FK constraints in current production systems~\cite{redset}. Since we envision \parachute{} columns being attached on the fly as the workload runs---e.g., building \parachute{} columns for those columns accessed with a relatively high number of distinct predicates---we propose the following solution: First, possible joins across the schema can be inferred from the workload itself as Parachute tracks possible parachute column placements. What remains is to determine whether we are dealing with a PK-FK relationship or not.

A lightweight check is to compute an approximation of the exact number of distinct values of both keys using a HLL sketch~\cite{hll}. To this end, note that, in principle, Parachute can easily handle \emph{inexact} PK-FK relationships if (a) a key appears in more than one tuple on the PK-side, or (b) a key on the FK-side has no associated key on the PK-side. Case (a) can be solved by \texttt{or}-ing out the column fingerprints coming from the tuples associated with the same key. In this case, the translation logic will require minor modifications. To accommodate case (b), \parachute{} columns can initially store a \texttt{NULL} value by default.  The problem arises when the matching tuple is eventually inserted on the PK-side. In this case, we would maintain a set of non-matching keys on the FK-tables and update them when the join condition is met.}

\sparagraph{Parachute-Aware Optimizer.} In its current state, Parachute acts on the ``frozen'' query plan, as output by the system's optimizer. This guarantees that we can maintain the efficient upwards information flow by \PSF{}, which also acts on the frozen query plan. In principle, it is possible to integrate parachute column estimates into the optimizer. \opt{Indeed, we can (over-)estimate the size of a PK-FK join by analyzing the translated parachute predicate on the FK-table, as follows:
\[
\begin{aligned}
& \left| \tabmk{}\:\join\:\sigma_{\texttt{production\_year}~<~2025}(\tabt{}) \right| \\
& \leq \left| \sigma_{\texttt{parachute\_title\_production\_year}~\leq~\texttt{bin}(2025)}(\tabmk{})\right|,
\end{aligned}
\]
where $\texttt{bin}(2025)$ is the corresponding bin of the constant 2025 in the attached parachute column (recall Fig.~\ref{fig:attach-vis}). Indeed, by using the statistics of the parachute column on \tabmk{}, we can \emph{overestimate} how many join partners we have from \tabt{} with the respective table predicate. While this is trivial for a single join, it is natural to ask how it develops with multiple joins, especially when integrating with \PSF{}; we leave this as an interesting future work. To estimate cardinalities on string fingerprints, we have to sum up the cardinalities of all supersets of the pattern fingerprint, \texttt{pmask}; formally, this is known as a (reversed) zeta transform~\cite{fsc}.}

\sparagraph{User Experience.} Currently, Parachute has a singl, $\pbw{}$. Note that each \parachute{} column, regardless of its data type, is reserved exactly \pbw{} bits (which may or may not be fully used, depending on its cardinality). However, as of now, there is no upper-bound on the ultimate extra space needed. Hence, an improvement in terms of usability is to replace \pbw{} with an \emph{upper-bound} on the extra space allowed, e.g., 10\%. The challenge is therefore to find optimal \pbw{}-values for each individual parachute column, while staying within the upper-bound and minimizing the false positive rate.

\sparagraph{Join-Induced Sorting.} Parachute can also be considered a data-driven variant of \MDDL{} on \emph{join-induced} filters. Hence, although primarily designed for query processing, \parachute{} columns can also be used for sorting (even jointly, i.e., considered together as a column by itself). A promising future work is to have hybrid, workload- \emph{and} data-driven parachute columns, i.e., have bits for repeating predicates, and otherwise the standard layout for columns with a large set of distinct predicates.

\sparagraph{Alternative Representations.} Another way of representing the implicit index induced by Parachute is to use row-ranges, as done in Amazon Redshift~\cite{pred_cache}. Concretely, we will still keep the histogram representation, yet each bin has a list of row-ranges that cover the tuples of that bin. To bound the space overhead (in case the row-ranges are not clustered), we can employ a similar strategy as in predicate caching~\cite{pred_cache}, to have an over-approximation of the row-ranges, at the cost of having more false positives.

%% file: sections/related-work.tex
\section{Related Work}\label{sec:related-work}

Given the recent (and understandable) focus of the database community, both research- \emph{and} industry-oriented, on robust query processing, we warn the reader that the related work will be rather verbose.
We will not simply enlist prior work, but we will also showcase how Parachute nicely fits into the respective context.
The section is split into the following subthemes: sideways-information passing (Sec.~\ref{subsec:rw-sip}), join indexes (Sec.~\ref{subsec:join-indices}), instance-optimal algorithms (Sec.~\ref{subsec:rw-ioa}),  and caching mechanisms (Sec.~\ref{subsec:pc}).

\subsection{Sideways-Information Passing}\label{subsec:rw-sip}

Sideways-information passing (SIP)~\cite{sip_1, sip_2} is a practical way to deal with the issue of expensive hash-tables probes, by building bloom filters on the build sides and pushing them into the upcoming pipelines. Across the paper, we argued that production systems implement a more restrictive form of SIP, namely probe-side filtering (PSF)~\cite{pred_cache}. However, there is another extension, which aims to optimize the order in which the pushed bloom filters are probed.

\sparagraph{LIP.} Lookahead-Information-Passing (LIP)~\cite{lip} makes the case that the order in which the pushed bloom filters are probed does indeed make a difference. Initially, LIP was designed to work on a star schema, yet it is now employed more generally within a single pipeline~\cite{lip_generalization} and has already been adopted in SQLite~\cite{sqllite}. Let us see a formalization of this paradigm for the case where only build sides push to the probe side of the same pipeline. We have:\[
    R \flow S \vcentcolon\Leftrightarrow (\pip(R) = \pip(S)) \land \texttt{is\_probe}(S),\]
since $R \leftrightarrow S$ holds by the definition of an execution pipeline. Hence, Parachute is also applicable when LIP is used. In addition, Zhang et al.~\cite{lip_generalization} describe yet another form of LIP, in which build sides can prune themselves. This still fits into our formalization, yet one needs to introduce a \texttt{rank}$(\cdot)$ function on the build sides of a pipeline to match their definition; we omit the details.

\subsection{Join Indexes}\label{subsec:join-indices}

Parachute can be regarded as a space-flexible bitmap join index~\cite{mt_joins_bji} or, more specifically, as a generalization of the dimension-join index~\cite{dimension-join}: The dimension-join index has a bitmap on the FK-table for each value of a low-cardinality dimension column. Parachute is a generalization of this index to (a) non-dimension tables---we can support \parachute{} columns coming from \tabt{} or \tabk{}, (b) high-cardinality columns: this is achieved via equi-depth histograms, which can be interpreted as a binning together the values in the dimension-join index, and (c) string support.

\subsection{Instance-Optimal Algorithms}\label{subsec:rw-ioa}

Instance-optimal algorithms also guarantee a bidirectional information flow by performing (approximate) semi-join reduction on the tables. As a consequence, they significantly reduce the number of dangling tuples. However, this strategy involves multiple passes over the data. Next, we outline several such approaches.

\sparagraph{Yannakakis' algorithm.} A prominent example is Yannakakis' algorithm~\cite{yannakakis}, which achieves full semi-join reduction, assuming an $\alpha$-acyclic query. It is composed of two phases: (a) the semi-join phase, and (b) the join phase. The full semi-join reduction is achieved in phase (a). In turn, it consists of a forward and a backward pass. In the forward pass, the algorithm traverses the join tree bottom-up, by performing a semi-join on the current table with its (already reduced) children. In the backward pass, the algorithm iterates the join tree in the reversed direction. Finally, it is guaranteed that the base tables have no dangling tuples.
Hence, the join phase can be executed using vanilla binary joins, as the intermediate sizes are guaranteed to not exceed their input's size.

\sparagraph{\mr{(Robust) Predicate Transfer.}} Admittedly, Yannakakis' algorithm has the expensive \emph{exact} semi-join phase. Indeed, having an \emph{approximate} semi-join reduction is enough. A first effort in this direction Predicate Transfer (PT)~\cite{pt}, which mimics Yannakakis' forward-backward information flow paradigm yet in an approximate sense, by resorting to bloom filters. Namely, PT introduces a transfer schedule of the bloom filters before the actual query plan is executed. \mr{However, despite its promising results, PT fails to provide robustness guarantees for $\alpha$-acyclic queries~\cite{rpt}. This issue was recently addressed in RPT~\cite{rpt}, which ensures a robust transfer schedule. Nevertheless, RPT still performs multiple passes over the data: the forward and backward passes of the bloom filters, and the (implicit) execution of the hash-joins. Parachute bypasses the need for the two additional passes by moving the overhead to storage.}

\sparagraph{Lookup \& Expand.} Another promising paradigm is Lookup \& Expand (L\&E) by Birler et al.~\cite{le}, and the later refinement by Bekkers~\cite{yannakakis_eng} for a more restricted class of $\alpha$-acyclic queries. The main idea is to split the hash-join operator into two intrinsically distinct operators, lookup and expand, that can be freely reordered in the query plan. L\&E may still benefit from \parachute{} columns in the case of non-$\alpha$-acyclic queries.

\subsection{Caching Mechanisms}\label{subsec:pc}

In recent work, it has been pointed that queries in real workloads are highly repetitive~\cite{redset, pred_cache}.
This led many systems rethink how they approach such queries.
Until recently, the state-of-the-art approach was a rather heavyweight materialization of query results, materialized views.
Redshift shifted towards considering a predicate cache~\cite{pred_cache} instead, where the first-class citizen is not the data itself anymore, as in the case of materialized views,
but a lightweight representation of the row indexes that are to be scanned. This makes updates much more easy-to-handle. The main primitive is a cache that holds table predicates and even pushed semi-join filters. This integrates well with Parachute since one can now also store the \emph{translated} parachute predicates. Since \parachute{} predicates come from PK-tables, which are less frequently updated than FK-tables, this increases the chance that they remain in the predicate cache.

%% file: sections/conclusion.tex
\section{Conclusion}\label{sec:conclusion}

In this work, we have made the case that current implementations of sideways information passing lack a bi-directional information flow, which instance-optimal algorithms de facto guarantee. In turn, the latter, including old proposals such as Yannakakis' algorithm~\cite{yannakakis}, and even more recent refinements~\cite{pt, rpt}, incur several passes over the data, which impedes their adoption in real systems. Motivated by the openness of production OLAP systems towards storing succinct auxiliary columns to improve query performance, e.g., MDDL~\cite{mddl}, we introduce Parachute as precomputed join-induced fingerprint columns that enable bi-directional information flow in a single pass. To achieve this, we propose several column representations for common data types, particularly focusing on string data, which is known to be the most frequent data type in data warehouses~\cite{redset}.

\sparagraph{Outlook.} Parachute has a strong advantage over dynamically pushed bloom filters, which makes it suitable to the cloud setting: Given that parachute columns are attached directly to the table, they can be leveraged during partition pruning. Moreover, Parachute can also blend well with recent caching paradigms, such as Predicate Caching~\cite{pred_cache}: Apart from caching semi-join filters, one can now also store the translated parachute predicates, enabling build-side pruning if the query plan contains a FK-table on the build side. In the face of workload changes where the cache has to be invalidated, we envision Parachute's data-driven design as a robust mechanism for dealing with workload changes.